\documentclass[journal=jacsat,manuscript=article]{achemso}

\usepackage[version=3]{mhchem} 

\author{Ermin Malic}
\email{ermin.malic@tu-berlin.de}
\affiliation{Fritz-Haber-Institut der Max-Planck-Gesellschaft, Faradayweg 4-6, D-14195 Berlin, Germany}
\alsoaffiliation{Institut f{\"u}r Theoretische
Physik,  Technische
Universit{\"a}t Berlin, 10623 Berlin, Germany}

\author{Heiko Appel}
\affiliation{Fritz-Haber-Institut der Max-Planck-Gesellschaft, Faradayweg 4-6, D-14195 Berlin, Germany}
\author{Oliver T. Hofmann}
\affiliation{Fritz-Haber-Institut der Max-Planck-Gesellschaft, Faradayweg 4-6, D-14195 Berlin, Germany}

\author{Angel Rubio}
\email{angel.rubio@ehu.es}
\affiliation{Nano-Bio Spectroscopy group and ETSF Scientific Development Centre, Universidad del Pais Vasco, Avenida de Tolosa 72, E-20018 Donostia, Spain}
\affiliation{Fritz-Haber-Institut der Max-Planck-Gesellschaft, Faradayweg 4-6, D-14195 Berlin, Germany}
\alsoaffiliation{Fritz-Haber-Institut der Max-Planck-Gesellschaft, Faradayweg 4-6, D-14195 Berlin, Germany}

\title
  {F{\"o}rster-induced energy transfer in functionalized graphene}

\abbreviations{IR,NMR,UV}

\begin{document}
\begin{abstract}

Carbon nanostructures are ideal substrates for functionalization with molecules, 
since they consist of a single atomic layer giving rise to an extraordinary sensitivity to changes in their surrounding. The functionalization opens a new research field of hybrid nanostructures with tailored properties.
Here, we present a microscopic view on the substrate-molecule interaction in the exemplary hybrid material consisting of 
graphene functionalized with perylene molecules. First experiments on similar systems have been recently realized illustrating an extremely efficient transfer of excitation energy from adsorbed molecules to the carbon substrate - 
a process with a large application potential for high-efficiency photovoltaic devices and 
biomedical imaging and sensing. 
So far, there has been  no microscopically founded explanation for the observed energy transfer.
Based on first-principle calculations, we have explicitly investigated the different transfer mechanisms revealing the crucial importance of 
F{\"o}rster coupling. Due to the efficient Coulomb interaction in graphene, we obtain strong F{\"o}rster rates in the range of 1/fs. We investigate its dependence on the substrate-molecule distance $R$ and describe the impact of the momentum transfer $q$ for an efficient energy transfer. Furthermore, we find that the Dexter transfer mechanism is negligibly small due to the vanishing overlap between the involved strongly localized orbital functions. 
The gained insights are applicable to a variety of carbon-based hybrid nanostructures.

\end{abstract}


The continuing trend to miniaturization of devices in modern technology leads to fundamental physical limits of applied materials.\cite{avouris07, ferrari10b} The search for new materials and new functionalities brings hybrid systems into the focus
of current research.\cite{hirsch05, burghard05} They consist of low-dimensional nanostructures functionalized with single molecules combining the remarkable properties
of  both subsystems.  In particular, carbon nanostructures are excellent substrates, since they offer a variety of metallic and 
semiconducting systems  showing a large sensitivity to changes in their surrounding.\cite{reichbuch, jorio08, erminbuch, avouris08b}  
Non-covalent functionalization based on $\pi-\pi$ stacking preserves the intrinsic properties of the substrate to a large extent.\cite{malic11} At the same 
time, the interaction with the attached molecules induces additional properties desired for specific technological applications.\cite{guo05,kern08, zhou09, hirsch09, charlier09, loi10, kolpak11}

First experiments have been realized illustrating the successful functionalization of carbon nanotubes with photoactive
molecules suggesting the design of efficient carbon-based molecular switching.\cite{guo05, simmons07, zhou09, malic12, setaro12} Recently, a strong excitation energy transfer has been observed in 
perylene- and  porphyrin-functionalized carbon nanotubes suggesting efficient photo-detection and light harvesting.\cite{voisin08,voisin10, ernst12} First studies 
on functionalized graphene also reveal high energy transfer rates  between the attached molecules and the graphene layer.\cite{koppens13} 
The combination of unique transport properties of graphene including ballistic transport and strong light absorption of organic molecules results in new hybrid nanostructures with large application potential for 
high-efficiency photodetectors, biomedical sensors, and photovoltaic devices.\cite{koppens13}\\

\begin{figure}[t!]
\center{\includegraphics[width=10cm]{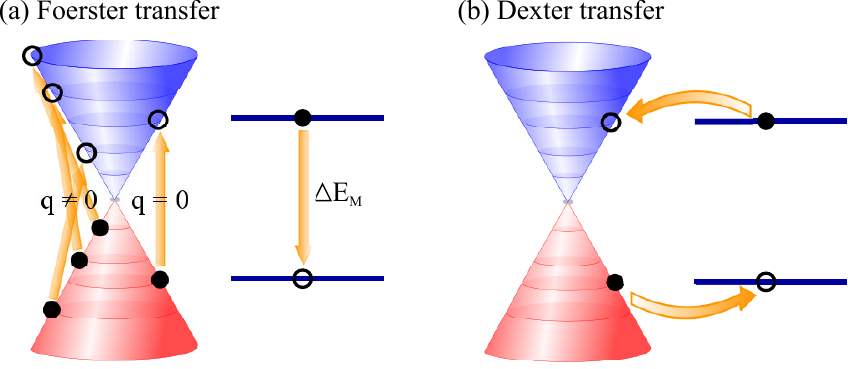}
 }
 \caption{ Schematic illustration of the non-radiative (a) F{\"o}rster\cite{foerster} and (b) Dexter\cite{dexter} energy transfer in perylene-functionalized graphene. The Dirac cone represents the electronic bandstructure of graphene, while the  electronic states of the perylene molecules are described by a two-level (HOMO-LUMO) momentum independent system. Different arrows in (a) show different energy-conserving processes involving varying momentum transfer $\mathbf{q}$. 
}\label{fig0}
\end{figure}

The observed energy transfer could be explained by two major non-radiative energy transfer mechanisms (as depicted in Fig. \ref{fig0}):\cite{winkler13} (i) F{\"o}rster coupling\cite{foerster}
describes a direct transfer of energy from the optically  excited molecule  to graphene. This leads to a quenching of the molecular emission, since the energy is non-radiatively transferred to the electrons in graphene, cf. Fig. \ref{fig0}(a). The F{\"o}rster transfer rate strongly depends on the molecular transition dipole moment $\mathbf{d}$ and it exhibits a $R^{-4}$ dependence for hybrid nanostructures on top of a spatially extended two-dimensional substrate\cite{swathi09, koppens13} (in contrast to the well-known $R^{-6}$ scaling for dipolar F{\"o}rster coupling in molecule-molecule complexes).  (ii) Dexter 
coupling\cite{dexter}  is based on a charge transfer between the molecule and graphene states, cf. Fig. \ref{fig0}(b). After the process, the molecule is brought into its ground state and graphene becomes excited and can emit light through carrier recombination. It is a short-range transfer mechanism 
that  directly depends on the spatial overlap of involved molecule and graphene orbital functions resulting in  an exponential decay with the substrate-molecule distance $R$.

Recent studies indicate that the observed energy transfer in carbon-based hybrid nanostructures can probably be traced back to a F{\"o}rster-like transfer process.\cite{koppens13, ernst13}
In these studies, the molecule-substrate distance is clearly larger than 10 \AA\, due to the presence of long non-conducting linker molecules. However, for functionalization procedures without such additional molecules, the distance is in the range of just a few \AA\, corresponding to the Van der Waals radius of the involved atoms.\cite{voisin08} Here, the Dexter transfer mechanism is expected to be a competing energy transfer mechanism. 
In this Letter, we present a systematic first-principle study on the substrate-molecule interaction in the exemplary hybrid system consisting of graphene functionalized with perylene molecules. The obtained insights should be applicable to other carbon-based hybrid nanostructures. We study the molecule-induced changes in the electronic bandstructure and the optical properties of graphene as well as the charge rearrangements within the two sub-systems. We explicitly calculate the F{\"o}rster transfer rate and investigate its importance as a function of the substrate-molecule distance $R$. Combining first-principle calculations with the tight-binding approximation, we obtain an analytic expression for the transfer rate. Furthermore, we discuss the competing Dexter transfer mechanism by estimating the spatial overlap of the involved substrate and molecule orbitals. 
 \\

\begin{figure}[t!]
\center{\includegraphics[width=10cm]{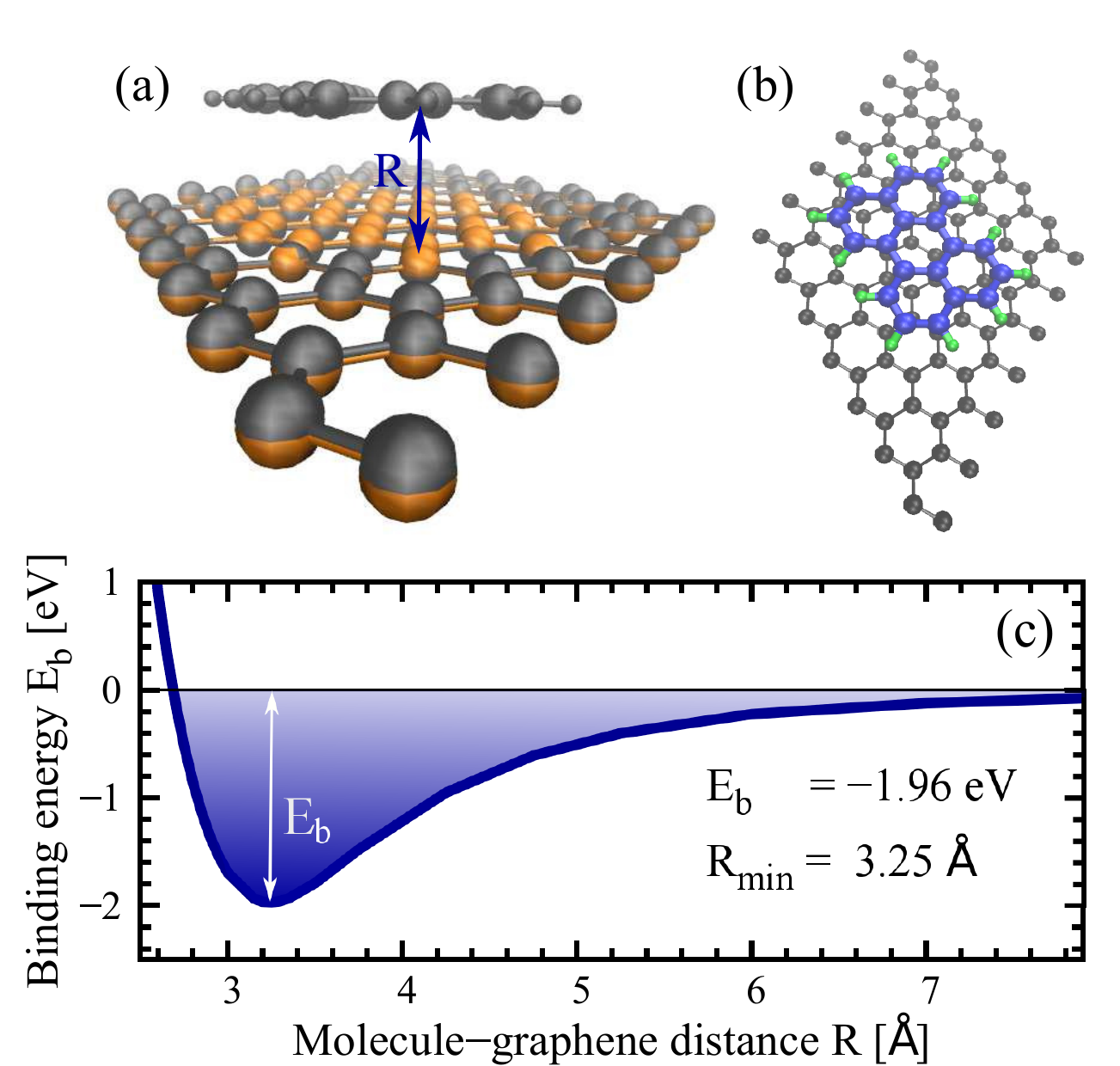}
 }
 \caption{(a) Perylene-functionalized graphene after full geometric relaxation within the FHI-aims code including the van der Waals interaction. For comparison, the initial position of carbon atoms within a perfectly flat graphene layer is shown in orange illustrating a slight dent of the graphene layer in the vicinity of the molecule after the geometric relaxation. (b) Top view on the relaxed hybrid nanostructure emphasizing the structure of the perylene molecule. (c) The binding energy $E_b$ as a function of the substrate-molecule distance $R$. 
}\label{fig1}
\end{figure}

The investigations are based on density
functional theory (DFT) calculations performed within the FHI-aims code package\cite{blum09}. It is an all-electron full-potential electronic structure code including numerical atom-centered orbitals, which are very efficient
allowing the investigation of structures containing hundreds of atoms. All calculations are done within the tight settings including a tier 2 basis set for the carbon and hydrogen atoms.\cite{blum09}. Calculations with increased accuracy in the basis functions revealed that the chosen settings already lead to converged results with respect to the total energy. We focus on graphene functionalized with perylene molecules ($C_{20}H_{12}$), cf. Figs. \ref{fig1}(a)-(b) illustrating the top and side view of the studied structure. For graphene, we choose a supercell covering $7x7$ unit cells corresponding to 98 carbon atoms with a lattice constant of 1.42 \AA.  The investigated situation corresponds to a moderate functionalization degree with a molecule-molecule distance of approximately 7 \AA. The electron interactions are described within the PBE exchange-correlation functional\cite{pbe96}  including the recently implemented Van der Waals 
correction\cite{tkatchenko09} to account for the long-rang van der Waals interaction. The latter plays a fundamental role in describing the weak molecule-nanostructure coupling that is of paramount importance to quantitative estimate the relative contribution of the Dexter transfer mechanism. We also performed additional calculations with the hybrid functional PBE0\cite{adamo99} and HSE06\cite{heyd03, krukau06}
to investigate the alignment of molecular levels.
We found that the molecular  HOMO (LUMO) level is located below (above) the Fermi energy in graphene and thus, initial spurious charge transfer does not occur. \\

The initial perylene-functionalized graphene structure is fully relaxed using the Broyden-Fletcher-Goldfarb-Shanno method minimizing all force components to values smaller than $10^{-3}$ eV/\AA. 
 Figure \ref{fig1}(a) illustrates the hybrid nanostructure after geometric relaxation. The comparison with the perfectly flat graphene layer (orange color) reveals a slight dent of carbon atoms of less than 0.1 \AA\, close to the molecule. This geometric pillow effect  is a direct consequence of the presence of the perylene molecule and can be traced back to the Pauli pushback\cite{bagus08,hofmann08}. It also gives rise to a charge rearrangement, which will be discussed below. 
 \begin{figure}[t!]
\center{\includegraphics[width=10cm]{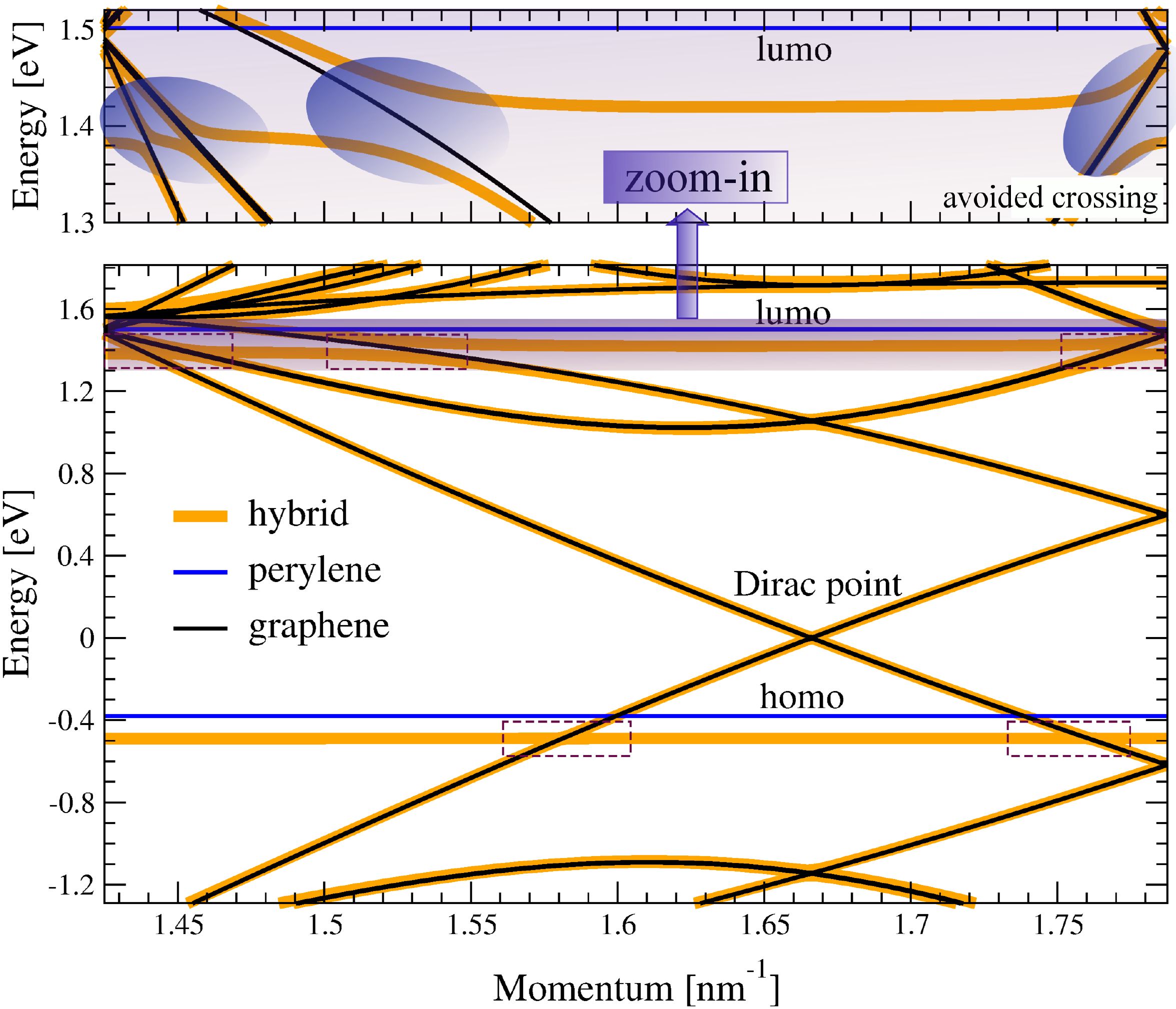}
 }
 \caption{Bottom panel: Electronic bandstructure of the hybrid nanostructure (orange lines) in direct comparison to the bandstructure of unfunctionalized graphene (black lines) and isolated perylene molecules (blue lines). The electronic states remain largely unchanged expect for the appearance of avoided crossings (dashed areas). Top panel: The region around the molecular LUMO level is zoomed-in to further illustrate this behavior.
}\label{fig2}
\end{figure}

We find an optimal substrate-molecule distance of $R_{min}=3.25$ \AA, which is slightly smaller than the initial value of the Van der Waals diameter of the carbon atom, cf. Fig. \ref{fig1}(c). The optimal binding energy at $R_{min}$ is $E_b=-1.96$ eV corresponding to a binding energy of  $E_b=-61$ meV per atom in the perylene molecule. This is in the expected range for a Van der Waals-induced non-covalent adsorption of the molecule to the graphene surface. 
The $\pi$-electronic system of the perylene molecule is linked to the graphene surface via $\pi-\pi$ stacking, which is much less invasive compared to the covalent adsorption.\cite{hirsch05} This can be well observed on the only minor changes in the electronic structure of the substrate, cf. Fig. \ref{fig2}. The unique bandstructure of graphene including the Dirac point and the linear bands is entirely preserved after the functionalization with perylene molecules. The observable changes appear at the points where the molecular HOMO and LUMO levels cross the graphene electronic states, as illustrated in the inset of Fig. \ref{fig2}. Here, the  resulting states of the hybrid nanostructure exhibit avoided crossings. This well-known behavior in quantum chemistry is further illustrated within the zoomed-in region around the molecular LUMO level, which anti-crosses the graphene electronic states several times.

\begin{figure}[t!]
\center{\includegraphics[width=10cm]{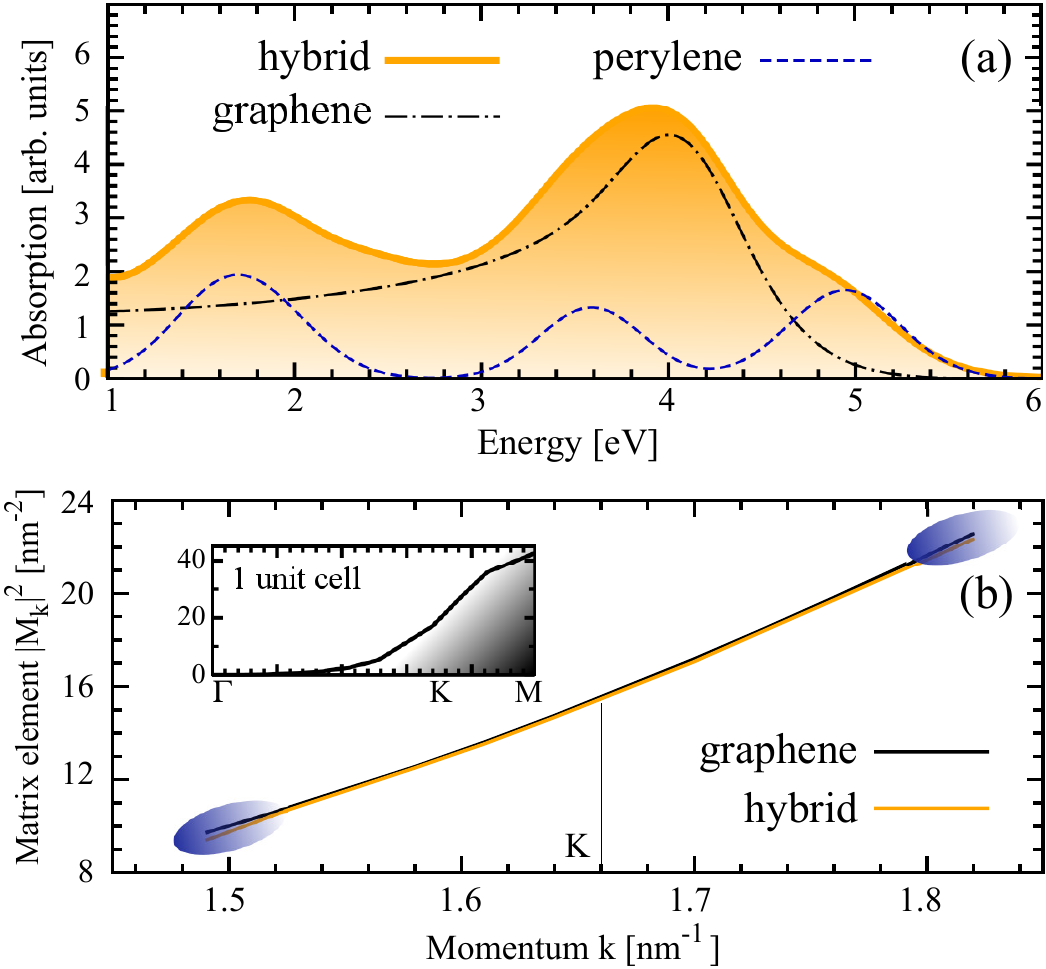}
 }
 \caption{(a) Optical absorption of the perylene-functionalized graphene in comparison to the absorption of the pristine graphene and the isolated perylene molecule.
 (b) Optical matrix element describing the strength of matter-light coupling for the hybrid nanostructure and pristine graphene, respectively. The inset shows the matrix element along the high-symmetry line $\varGamma$KM within the Brillouin zone of one-unit-cell graphene.
}\label{fig3}
\end{figure}

As a direct consequence of the almost completely preserved electronic bandstructure, the optical properties of graphene remain unchanged to a large extent, cf. Fig. \ref{fig3}. The optical absorption of the hybrid nanostructure corresponds to an overlap of the absorption peaks of the pristine graphene and isolated perylene molecule, as shown in Fig. \ref{fig3}(a). The optical matrix element corresponding to the expectation value of the momentum operator\cite{erminbuch} $\mathbf{M^{vc}_k}=\langle \psi^v_k| \mathbf{p}| \psi_k^c \rangle$ exhibits only slight changes in the region, where avoided crossing takes place, cf. the blue-shaded circles in Fig. \ref{fig3}(b).  Note, however, that the energy transfer within the hybrid nanostructure is not directly included within the DFT treatment and will be further discussed below.

The absorption spectrum of graphene is characterized by the well pronounced peak at approximately 4 eV corresponding to the transition at the saddle point (M point) in the Brillouin zone.\cite{heinz11, giessen11, malic11} The widely delocalized $\pi$ electronic system in the perylene molecule gives rise to strong absorption peaks at 1.7 eV, 3.6 eV, and 4.9 eV. The obtained transition energies are lower than in experiment due to the shortcoming of the applied exchange-correlation functional. Calculations based on hybrid functionals give a much better agreement with the experiment. Since in this work, we focus on the F{\"o}rster and Dexter energy transfer mechanisms between the perylene molecule and graphene, the energetic deviations within PBE exchange-correlation function do not play a qualitative role. Due to the linear gap-less bandstructure of graphene in the relevant energy region, there are always electronic states that are in resonance with the energetically lowest HOMO-LUMO transition of the perylene molecule.

\begin{figure}[t!]
\center{\includegraphics[width=10cm]{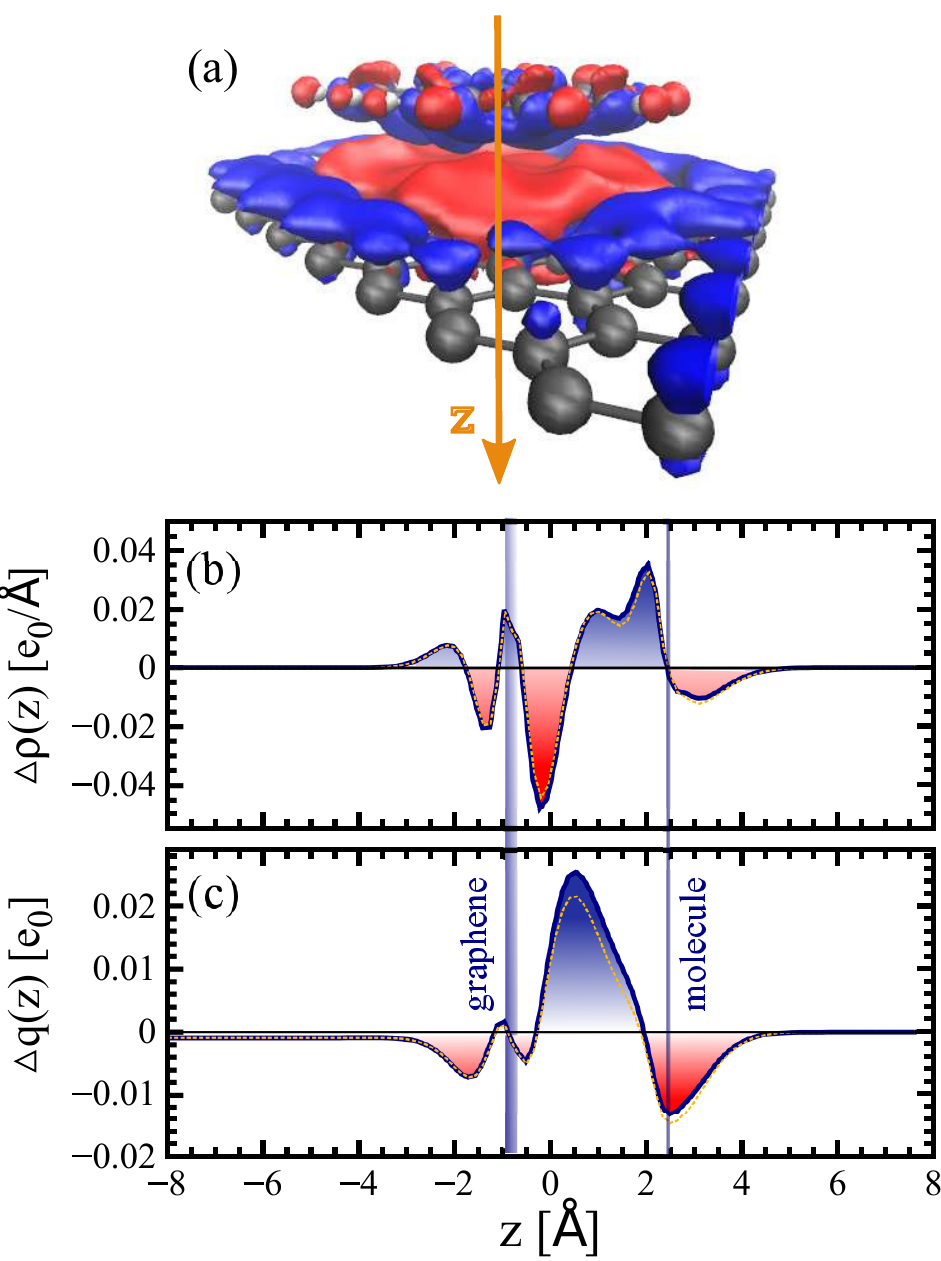}
 }
 \caption{The charge density difference $\Delta \rho(x,y,z)$ is illustrated within a surface plot for the exemplary iso-value of $\pm 5\times 10^{-4}\, e_0 /$\AA$^3$. The red color corresponds to the negative value describing the lack of electrons, while the blue color reflects electron accumulations. (b) Charge density difference $\Delta \rho(z)$ integrated over the xy-plane illustrating the change of $\Delta \rho$ along the z-axis perpendicular to the graphene layer. (c) Charge difference $\Delta q(z)=\int^z_{-\infty} dz' \Delta \rho (z')$ accumulated along the z-axis as a function of z. The dashed lines correspond to the result obtained within constrained DFT calculations. 
}\label{fig4}
\end{figure}

Furthermore, we have investigated the charge rearrangement within the hybrid nanostructure. As already seen in Fig. \ref{fig1}, the adsorbed molecule leads to a spatial pillow-like effect\cite{hofmann13} pushing the graphene's carbon atoms further away and giving rise to a small dent of  < 0.1 \AA\,. This also affects the mobility of charge carriers within the graphene layer resulting in charge rearrangements. Figure \ref{fig4}(a) shows an surface plot illustrating the molecule-induced charge density difference $\Delta \rho (x,y.z)=\rho_{\text{hybrid}}-\rho_{\text{graphene}} -\rho_{\text{perylene}}$ for the exemplary iso-value of $5\times 10^{-4} \, e_0 / $\AA$^3$. One can clearly see the
accumulation of negative (blue) and positive (red) charges.
According to the pillow effect, the electrons are pushed away from the region directly below the molecule. As a result, this region is characterized by a positive charge, i.e. the lack of electrons (red areas). At the same time, electrons accumulate further away  at the graphene surface at the graphene-facing side of the molecule (blue areas). 
To further illustrate the charge distribution along the z-direction (perpendicular to the graphene surface), we show the plane-averaged charge density difference $\Delta \rho (z)=\int dx \int dy \Delta \rho (z)$  and the charge difference $ \Delta q(z)=\int^z_{-\infty} dz' \Delta \rho (z')$, cf. Figs. \ref{fig4}(b) and (c), respectively. The charge distribution around the graphene layer qualitatively reflects the spatial shape of the most relevant $2p_z$ carbon orbitals reaching above and below the graphene sheet. A similar charge distribution can also be observed around the position of the perylene molecule illustrating a positive (negative) charge accumulation slightly below (above) the molecule. 
Note, however, that the quantitative effect of charge rearrangements is relatively small. The predicted small charge difference of up to $0.02\,e_0$ is in agreement with what one would expect for a non-covalent functionalization.

The dashed lines in Fig. \ref{fig4} reflect the charge distribution obtained within constrained DFT calculations,\cite{voorhis05} i.e. we imposed particular initial occupations of molecular HOMO and LUMO levels while solving the Kohn-Sham equations. The aim was to investigate the change of the substrate-molecule interaction once the molecule is optically excited. Therefore, we promoted one electron from the HOMO to the LUMO level. The calculations show only marginal changes in the charge distribution (cf. Fig. \ref{fig4}) or in the electronic bandstructure (not shown). This insight is important for the discussion of the excitation energy transfer in the investigated hybrid structure. Time-dependent DFT calculations\cite{kuemmel13} are beyond the scope of this study and will be performed in future work.
\\

After having characterized the perylene-functionalized graphene including its electronic and optical properties, we now focus on the investigation of the possible energy transfer mechanisms in such a hybrid structure, cf. Fig. \ref{fig0}. The F{\"o}rster and Dexter energy transfer rates can be analytically expressed via the Fermi golden rule
\begin{equation}
 \label{gammaF}
\gamma =\dfrac{2\pi}{\hbar}\sum_{\mathbf{k_i}} \sum_{\mathbf{k_f}} \left|V(v\mathbf{k_i},c\mathbf{k_f}, l, h) \right|^2 \delta(\varepsilon_\mathbf{k_f}^c-\varepsilon_\mathbf{k_i}^v-\Delta E_M)
\end{equation}
with the momentum-dependent initial and final states of the graphene substrate $\Phi^{\lambda}_\mathbf{k_i/k_f}(\mathbf{r})$ and the HOMO and LUMO states of the molecule $\Phi^{l/h}_M(\mathbf{r})$. The delta function makes sure that only energy-conserving processes contribute. The F{\"o}rster rate $\gamma_F$ is determined by the direct contribution of the Coulomb interaction\cite{foerster}
 \begin{equation}
 \label{foerster}
V_F(v\mathbf{k_i},c\mathbf{k_f}, l, h)=\dfrac{e_0^2}{4\pi\epsilon_0} \int d\mathbf{r}\int  d\mathbf{r'} \Phi^{v*}_\mathbf{k_i}(\mathbf{r}) \Phi^{c}_\mathbf{k_f}(\mathbf{r})
 \dfrac{1}{|\mathbf{r}-\mathbf{r'}|}  \Phi^{l*}_M(\mathbf{r'})   \Phi^{h}_M(\mathbf{r'}) ,
 \end{equation}
where $e_0$ denotes the elementary charge and $\epsilon_0$ the vacuum permitivity.
The exchange Coulomb contribution gives the Dexter rate $\gamma_D$ with\cite{dexter}
 \begin{equation}
 \label{dexter}
V_D(v\mathbf{k_i}, c\mathbf{k_f}, l, h)=\dfrac{e_0^2}{4\pi\epsilon_0} \int d\mathbf{r}\int  d\mathbf{r'} \Phi^{v*}_\mathbf{k_i}(\mathbf{r}) \Phi^{h}_M(\mathbf{r}) 
 \dfrac{1}{|\mathbf{r}-\mathbf{r'}|} \Phi^{l*}_M(\mathbf{r'}) \Phi^c_\mathbf{k_f}(\mathbf{r'})  . 
 \end{equation}
 For Dexter coupling, a large spatial overlap between graphene and molecular orbitals ($\Phi^{v*}_\mathbf{k_i}(\mathbf{r}) \Phi^{h}_M(\mathbf{r})$ and $\Phi^{l*}_M(\mathbf{r'})\Phi^c_\mathbf{k_f}(\mathbf{r'}) $) is of key importance.\cite{dexter, winkler13} As a result,  $\gamma_D$ shows an exponential dependence on the substrate-molecule distance $R$ and  occurs only for small distances (typically, smaller than 10 \AA).\cite{winkler13} 
In contrast, the F{\"o}rster coupling is dominated by the factor $\frac{1}{|\mathbf{r}-\mathbf{r'}|}$ in Eq. (\ref{foerster}). 

Considering the conventional energy transfer between donor and acceptor molecules, the F{\"o}rster coupling is based on the dipole-dipole interaction and is characterized by a $R^{-6}$ dependence.\cite{foerster, baer08, winkler13} 
In the case of functionalized graphene, the substrate is not a spatially localized molecule, but a periodically extended two-dimensional nanostructure.   Following the approach of Swathi et al.\cite{swathi09}, the F{\"o}rster coupling can be considered as an interaction of the molecular transition dipole 
  $
\mathbf{d_M}=-e_0\int d\mathbf{r'} \Phi^{l*}_M(\mathbf{r'}) \mathbf{r'} \Phi^{h}_M(\mathbf{r'}) 
 $
located in the electrostatic potential 
$
\varphi^{vc}_{\mathbf{k_i},\mathbf{k_f}}(\mathbf{r'})=\frac{1}{4\pi\epsilon_0}\int d\mathbf{r} \frac{\rho^{vc}_{\mathbf{k_i},\mathbf{k_f}}(\mathbf{r})}{|\mathbf{r}-\mathbf{r'}|} 
 $
arising from the transition charge density of graphene 
$
\rho^{vc}_{\mathbf{k_i},\mathbf{k_f}}(\mathbf{r})=-e_0 \Phi^{v*}_\mathbf{k_i}(\mathbf{r})  \Phi^{c}_\mathbf{k_f}(\mathbf{r}) .
$
Then, the  F{\"o}rster energy transfer rate can be written as
\begin{equation}
 \label{gammaF2}
\gamma_F =\dfrac{2\pi}{\hbar}\sum_{\mathbf{k_i}} \sum_{\mathbf{k_f}} \left| \mathbf{d_M}\cdot \nabla \varphi^{vc}_{\mathbf{k_i},\mathbf{k_f}}    \right|^2 \delta(\varepsilon_\mathbf{k_f}^c-\varepsilon_\mathbf{k_i}^v-\Delta E_M),
\end{equation}
where the electrostatic potential $\varphi^{vc}_{\mathbf{k_i},\mathbf{k_f}}$ is evaluated at the fixed position of the molecule.

Combining DFT calculations on the molecular transition dipole moment with the tight-binding approximation of the graphene wave functions allows us to obtain an analytic expression for $\gamma_F$. For the molecular transition dipole moment, we obtain
$\mathbf{d_M}=(d_x,d_y,d_z)=(-0.80,1.39,4.92x10^{-5})$ e$_0$\AA\, with $d_M=1.60$ e$_0$\AA. As expected for the flat perylene molecule lying in the x-y plane, $d_z$ is nearly zero. The dipole moment  is obtained  for the perylene molecule that has been fully geometrically relaxed in the presence of the graphene substrate. Furthermore, we have also performed constrained DFT calculations\cite{voorhis05} modeling an initially excited molecule (one electron promoted from the HOMO into the LUMO level) to account for the changes of the molecular states due to the optical excitation taking place before the actual energy transfer process, as illustrated in Fig. \ref{fig0}. Our calculations reveal only negligibly small changes of the dipole components in agreement with the marginal changes observed for the charge distribution in Fig. \ref{fig4}.
\begin{figure}[t!]
\center{\includegraphics[width=12cm]{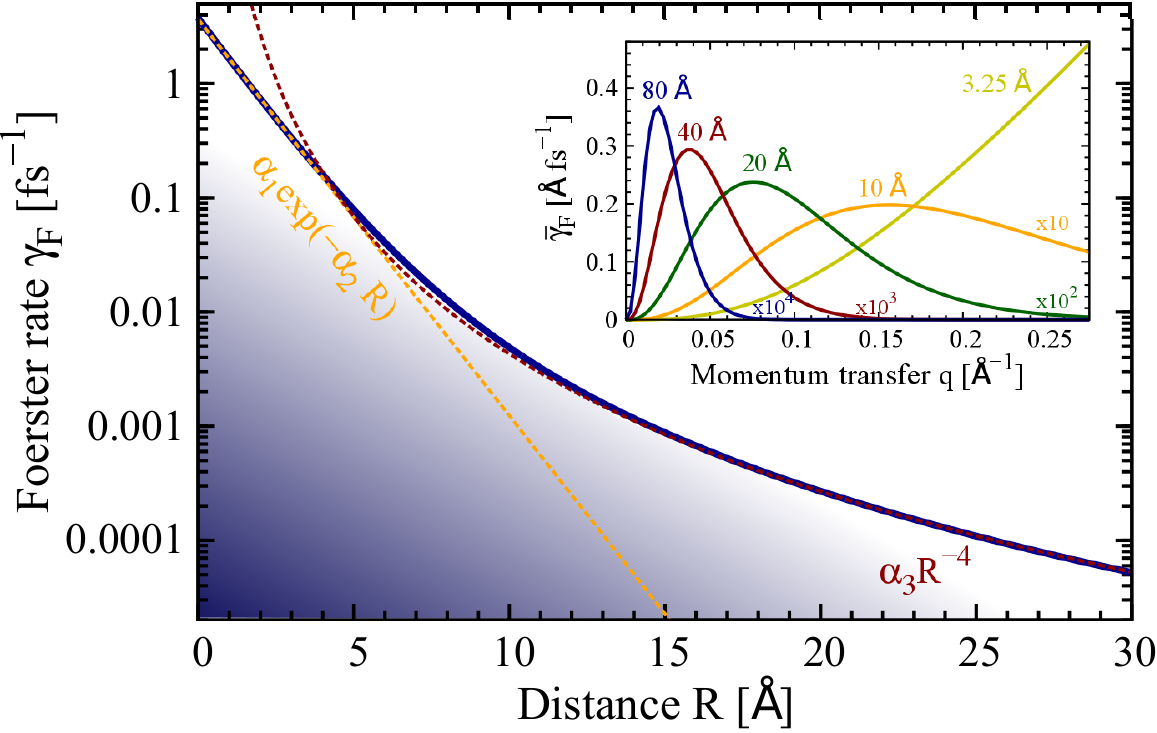}
 }
 \caption{F{\"o}rster energy transfer rate $\gamma_F$ as a function of the substrate-molecule distance $R$. For short distances, the transfer rate is characterized by an exponential decay (orange line), while for large distances a $R^{-4}$ behavior (red line) is found. The inset shows the momentum dependence of the processes contributing to the F{\"o}rster transfer rate at different constant substrate-molecule distances $R$. Here, $\overline{\gamma}_F$ corresponds to the integrand of Eq. (\ref{foersterF3}). 
}\label{fig5}
\end{figure}

Within the tight-binding approximation, the transition charge density of graphene 
$
\rho^{vc}_{\mathbf{k_i},\mathbf{k_f}}(\mathbf{r})$ can be obtained analytically. Taking into account only the strongest overlaps one obtains for the F{\"o}rster transfer rate\cite{swathi09}
 \begin{equation}
\label{foersterF3}
\gamma_F (R)=\int_0^{\frac{\Delta E_M}{\nu_F}} \dfrac{e_0^2}{128\pi\hbar\epsilon_0^2}(d_{\parallel}^2+2d_{\perp}^2) \dfrac{e^{-2qR}q^3}{\sqrt{\Delta E_M^2-\nu_F^2 q^2}}dq\;\approx\; \begin{cases} \alpha_1e^{-\alpha_2 R} & \mbox{for ~R\,\,\,<\,\,\,10 \AA} \\ \alpha_3R^{-4} & \mbox{for ~ R \,$\gg$ 10 \AA} \end{cases}\,
\end{equation}
with the HOMO-LUMO gap $\Delta E_M$ and the slope in the electronic bandstructure of graphene $\nu_F$.
 The F{\"o}rster coupling explicitly depends on the square of the parallel $d_\parallel$ (in the x-y plane)  and the perpendicular  component $d_\perp$ (z-axis) of the molecular transition dipole moment $\mathbf{d_M}$. For the investigated perylene-functionalized graphene,  $d_\perp$ can be neglected, as shown above.
 
Figure \ref{fig5} illustrates the F{\"o}rster rate as a function of the substrate-molecule distance $R$. 
Generally, the integral over all processes involving the momentum transfer $q$ cannot be analytically solved. We find that within the simplest tight-binding approximation taking into account only the strongest overlaps, the direct transitions with $q=0$ do not contribute to the energy transfer. The inset of Fig. \ref{fig5} shows the integrand of Eq. (\ref{foersterF3}) (denoted as $\overline{\gamma}_F$) as a function of $q$ for different fixed distances $R$. For R<10 \AA,  $\overline{\gamma}_F(q)$ quickly increases with $q$ and the F{\"o}rster rate  $\gamma_F$ shows an exponential decay with  $R$, i.e. $\gamma_F\approx\alpha_1e^{-\alpha_2 R}$ with $\alpha_1\approx3.66$ fs$^{-1}$ and $\alpha_2\approx 0.80$ \AA$^{-1}$, cf. Fig. \ref{fig5}.  For large distances, the behavior drastically changes: $\overline{\gamma}_F(q)$ is characterized by a maximum centered at $q\approx \frac{3}{2R}$, i.e. only processes involving a certain momentum transfer $q$ significantly contribute to the energy transfer rate. 
In the limit of large substrate-molecule distances ($R\gg10$ \AA), the F{\"o}rster coupling exhibits a clear $R^{-4}$ dependence, i.e. $\gamma_F\approx\alpha_3R^{-4}$ with $\alpha_3\approx42.85$ fs$^{-1}$, cf. Fig. \ref{fig5}. This is in excellent agreement with the observations in a recent experiment varying the distance between graphene and attached molecular emitters by depositing additional layers.\cite{koppens13}

Inserting the molecular transition dipole moment $\mathbf(d_M)=(d_\perp, d_\parallel)$ for the investigated exemplary perylene-functionalized graphene, we obtain a very efficient F{\"o}rster energy transfer rate of $\gamma_F (R_0)=0.277$ fs$^{-1}$.  This can be traced back to the strong Coulomb interaction in  the graphene substrate and the short substrate-molecule distance of $R_0=3.25$ \AA\, obtained within a full geometric relaxation of the entire hybrid nanostructure. At such a short distance,  transitions involving different momentum transfers $q$ crucially contribute to the F{\"o}rster rate, cf. the inset of Fig. \ref{fig5}.  
Our result is in  line with experimental time-resolved investigations of the energy transfer in functionalized carbon nanotubes suggesting 
that the transfer process occurs on an ultrafast femtosecond timescale.\cite{voisin11} 
Often, it is necessary to include additional linker molecules to experimentally achieve the functionalization\cite{setaro12} resulting in much larger substrate-molecule distances. For example,  $R=10$ \AA\, and 50 \AA\, result in a F{\"o}rster rate of $\gamma_F =4.88$ ps$^{-1}$ and 6.88 $\times 10^{-3}$ ps$^{-1}$, respectively. The drastic decrease in efficiency is in agreement with the experimental findings of  L. Gaudreau and co-workers.\cite{koppens13}
\\

In spite of the short distance between the graphene layer and the perylene molecule, our calculations reveal that the Dexter energy transfer rate $\gamma_D$ is negligibly small compared to the discussed F{\"o}rster transfer mechanism. The Dexter rate is determined by the spatial overlap between the strongly localized graphene and perylene orbitals. To estimate $\gamma_D$, we calculate the ratio between the overlaps $\alpha_D=\langle \Phi^{v*}_\mathbf{k_i}(\mathbf{r})| \Phi^{h}_M(\mathbf{r})\rangle$ and $\alpha_F=\langle \Phi^{v*}_\mathbf{k_i}(\mathbf{r})| \Phi^{c}_\mathbf{k_f}(\mathbf{r})\rangle$ appearing in the Dexter and the F{\"o}rster rate, respectively, cf. Eqs. (\ref{dexter}) and (\ref{foerster}). We obtain  $\alpha_D/\alpha_F\approx 10^{-1}$. Since in the rates the square of the product of two such overlaps appears, the Dexter rate $\gamma_D$ is  
 expected to be approximately four orders of magnitude smaller than the F{\"o}rster rate $\gamma_F$. 
\\

In conclusion, we have investigated the energy transfer in perylene-functionalized graphene.  Having characterized the hybrid material within DFT calculations including a fully geometric relaxation of the structure, its electronic bandstructure, optical properties, and charge rearrangements, we focus on the energy transfer that has been measured in recent experiments.  Combining DFT-based calculation of the molecular transition dipole moment and tight-binding-based consideration of graphene wave functions allows us to obtain an analytic expression for the F{\"o}rster energy transfer rate. Our calculations reveal strongly efficient F{\"o}rster coupling with rates in the range of fs$^{-1}$. In contrast, the Dexter energy transfer mechanism is found to be negligibly small due to small overlap between the involved strongly localized substrate and molecule orbital functions. The obtained results can be applied to other carbon-based hybrid nanostructures and in general to the description of energy transfer processes in molecular functionalised nanostructures, once the molecular dipole moment and the substrate-molecule separation are known.\\

We thank the Einstein Foundation Berlin and the Deutsche Forschungsgemeinschaft (within the collaborative research center SFB 658) for financial support. O. T. H. acknowledges the support from FWF (Project: J 3285-N20). A.R acknowledge financial support from the European Research Council 
(ERC-2010-AdG-267374), Spanish Grant (FIS2010-21282-C02-01),
Grupos Consolidados UPV/EHU del Gobierno Vasco (IT578-13), and the  EU project  (280879-2 CRONOS CP-FP7) .






\providecommand*\mcitethebibliography{\thebibliography}
\csname @ifundefined\endcsname{endmcitethebibliography}
  {\let\endmcitethebibliography\endthebibliography}{}

\end{document}